\documentclass[sigconf,natbib=true,anonymous=false]{acmart}

\usepackage{balance} 
\usepackage{enumitem}

\AtBeginDocument{%
  }

\setcopyright{acmlicensed}

\copyrightyear{2026}
\acmYear{2026}
\setcopyright{cc}
\setcctype{by}
\acmConference[CHIIR '26]{2026 ACM SIGIR Conference on Human Information Interaction and Retrieval}{March 22--26, 2026}{Seattle, WA, USA}
\acmBooktitle{2026 ACM SIGIR Conference on Human Information Interaction and Retrieval (CHIIR '26), March 22--26, 2026, Seattle, WA, USA}
\acmPrice{}
\acmDOI{10.1145/3786304.3787866}
\acmISBN{979-8-4007-2414-5/2026/03}

\begin{document}

\title{"Can You Tell Me?": Designing Copilots to Support Human Judgement in Online Information Seeking}

\author{Markus Bink}
\affiliation{
  \institution{Neu-Ulm University of Applied Sciences}
  \city{Neu-Ulm}
  \country{Germany}}
\affiliation{
  \institution{University of Regensburg}
  \city{Regensburg}
  \country{Germany}  
  }
\email{markus.bink@ur.de}

\author{Marten Risius}
\affiliation{
  \institution{Neu-Ulm University of Applied Sciences}
  \city{Neu-Ulm}
  \country{Germany}}
\email{marten.risius@hnu.de}

\author{Udo Kruschwitz}
\affiliation{
  \institution{University of Regensburg}
  \city{Regensburg}
  \country{Germany}}
\email{udo.kruschwitz@ur.de}

\author{David Elsweiler}
\affiliation{
  \institution{University of Regensburg}
  \city{Regensburg}
  \country{Germany}}
\email{david.elsweiler@ur.de}

\renewcommand{\shortauthors}{Bink et al.}

\begin{abstract}

Generative AI (GenAI) tools are transforming information seeking, but their fluent, authoritative responses risk overreliance and discourage independent verification and reasoning. Rather than replacing the cognitive work of users, GenAI systems should be designed to support and scaffold it. Therefore, this paper introduces an LLM-based conversational copilot designed to scaffold information evaluation rather than provide answers and foster digital literacy skills. In a pre-registered, randomised controlled trial (N=261) examining three interface conditions including a chat-based copilot, our mixed-methods analysis reveals that users engaged deeply with the copilot, demonstrating metacognitive reflection. However, the copilot did not significantly improve answer correctness or search engagement, largely due to a \textit{"time-on-chat vs. exploration"} trade-off and users’ bias toward positive information. Qualitative findings reveal tension between the copilot’s Socratic approach and users’ desire for efficiency. These results highlight both the promise and pitfalls of pedagogical copilots, and we outline design pathways to reconcile literacy goals with efficiency demands.

\end{abstract}





\maketitle

\section{Introduction}
Online platforms and social media networks have become central channels for the rapid spread of misinformation, affecting topics from climate change \cite{zhou2022confirmation} and politics \cite{prochaska2023mobilizing} to health \cite{do2022infodemics}. Such claims are often difficult to verify, particularly when presented out of context \cite{newman2024misinformed, fazio2020out}. Susceptibility is further influenced by individual traits, including lower analytical thinking \cite{bronstein2019belief}, higher extraversion \cite{calvillo2024personality}, and specific political leanings \cite{guess2019less}. This is largely due to people's lack of adequate information literacy \cite{guess2020digital, stauch2025digital}, an effect amplified by people often overestimating their abilities \cite{mahmood2016people}, underscoring the critical need for effective strategies to help users evaluate the truthfulness of online content \cite{aslett2024online, elsweiler2025query}.

Traditionally, users have relied on search engines to explore the search space and evaluate claims they read online \cite{aslett2024online, elsweiler2025query}. However, the emergence of Generative AI (GenAI) has opened up new avenues in the information-seeking domain \cite{white2024advancing}. These systems, powered by Large Language Models (LLMs), are not only able to answer users' information needs directly \cite{white2024advancing}, a quality that users have come to expect \cite{stamou2010interpreting}, they are also capable of summarising large bodies of text \cite{zhang2024comprehensive} and adapt it to different literacy levels \cite{freyer2024easy}. 

While these features offer convenience, they also create a fundamental dilemma: users increasingly desire fast, fluent answers, yet these very answers risk bypassing the critical evaluation processes necessary to distinguish accurate from misleading information. Because LLMs are trained on large-scale data, their outputs inherit biases from this data \cite{wan2023kelly, dai2024bias, BaezaYates18Bias, Alonso24Information}. The apparent perfection of their responses can lead users to accept inaccuracies as fact \cite{mayerhofer2025blending}, while also diminishing their tendency to evaluate output critically \cite{kosmyna2025your}. Moreover, directly presenting users with answers reduces interaction \cite{bink2022featured, williams2016detecting}, which can shape reasoning based on only a brief snippet of text \cite{bink2023investigating}. Some even view GenAI systems as potential replacements for search engines altogether \cite{haque2022think}. These risks are compounded by the fact that people often overestimate their information literacy abilities \cite{mahmood2016people}, underscoring the critical need for systems that support effective strategies for evaluating truthfulness online \cite{aslett2024online, elsweiler2025query}.

Yet paradoxically, the very features that make LLMs risky, like their fluency and ease of use, also make them promising for such support. To address this, we draw on constructivist learning theory to frame our copilot as a support for active, self-directed knowledge construction. In addition, the copilot incorporates principles of cognitive scaffolding \cite{van2002scaffolding}, particularly by modeling expert reasoning strategies such as lateral reading and click restraint. The purpose is not to teach content per se, but to foster metacognitive skills for evaluating information online. This approach is supported by evidence that GenAI systems can aid learning and improve educational outcomes, from making driver education more engaging \cite{murtaza2024transforming} to serving as virtual tutors \cite{lo2023impact} and adapting text to learners’ literacy levels \cite{murgia2023children}.

Therefore, in this work, we propose an LLM-based conversational copilot grounded in digital literacy principles. The system is designed to aid users in decision-making and information evaluation by offering strategies that empower them to independently verify or question the claims they encounter online. Crucially, the copilot avoids providing direct answers, instead focusing on enhancing users' competences \cite{herzog2025boosting, bink2024balancing}.

We address the following research questions:
\begin{itemize}
    \item \textbf{RQ1:} How does the presence of a copilot influence users' search outcomes?

    \item \textbf{RQ2:} How do copilot responses influence user interaction with search engine result pages (SERPs)? 

    \item \textbf{RQ3:} How do users interact with a copilot aimed at teaching digital literacy skills? 

    \item \textbf{RQ4:} What do users expect from a copilot aimed at teaching digital literacy skills? 
\end{itemize}
\begin{figure*}
  \frame{\includegraphics[width=\textwidth]{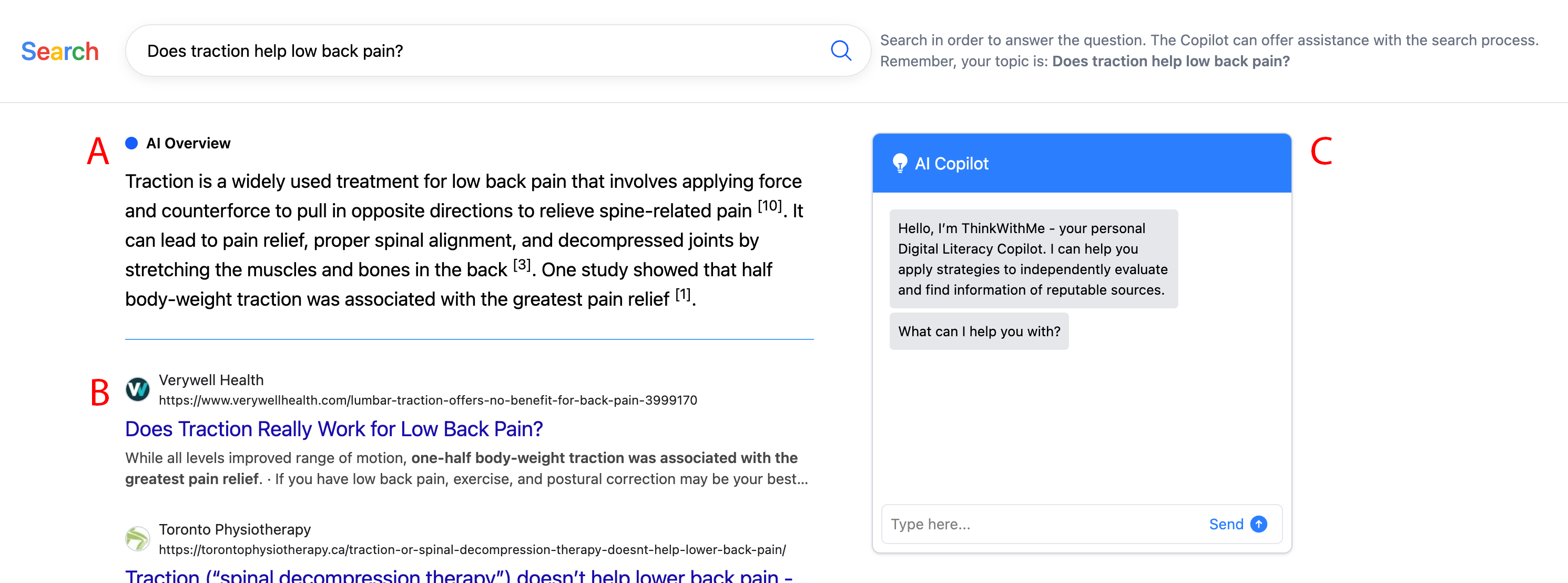}}
  \caption{Example SERP with search results on the left and the proposed interactive copilot on the right. A) Depicts the generated AI Overview for the query "Does traction help low back pain?", B) Depicts 10 search results, C) Depicts the copilot}
  \label{fig:serp}
\end{figure*}

\section{Related Work}
Our work builds upon several areas of research. First, we examine how users interact with search results and the biases that influence their decision-making. Second, we discuss interventions from behavioural science, specifically nudging and boosting, as strategies to shape information use. Third, we review existing work on generative AI and its application in search contexts. Finally, we cover literature regarding information literacy concepts.

\subsection{Biases in Web Search}

Biases in web search arise from how systems rank content, how results are presented, and how users process information \cite{BaezaYates18Bias}. Together, these factors illustrate that search engines do not provide a neutral window onto information, but instead shape what perspectives are encountered and how they are interpreted.

One strand of work shows that ranking and retrieval mechanisms themselves embed systematic biases. Rather than offering an even distribution of perspectives, search engines tend to privilege already popular or top-ranked items \cite{clarke2007influence,pandey2004wic}. This dynamic often amplifies dominant viewpoints \cite{white2013beliefs,white2014belief} and narrows the range of perspectives users encounter, even when queries are phrased neutrally \cite{gao2020toward,draws2023viewpoint}. Such findings highlight that the very architecture of search can subtly steer users toward particular interpretations.  

Equally influential is the way results are presented. The design of snippets, URLs, and other visual cues directly shapes user attention and click behaviour \cite{clarke2007influence, schwarz2011augmenting,yamamoto2011enhancing}. Search results that provide a direct-answer to users' information needs increase satisfaction and capture disproportionate attention \cite{wu2020directanswers}, yet they also discourage broader exploration of the SERP, showcasing its influence in how opinions are formed \cite{bink2022featured,bink2023investigating, wong2025retrieval, narayanan2025search}.

A further source of bias stems from users’ own cognitive tendencies. Attention is disproportionately drawn to top-ranked results \cite{joachims2007evaluating}, and repeated exposure to certain perspectives increases the likelihood of adopting them \cite{draws2021sigir}. Such dynamics underpin the Search Engine Manipulation Effect (SEME), where prominent viewpoints can shift attitudes, particularly among users without strong prior opinions, often without their awareness \cite{allam2014impact,bink2023investigating,draws2021sigir,epstein2015seme}. Finally, users also tend to favour positive over negative information \cite{white2013beliefs}.  

\subsection{Nudging \& Boosting}

\emph{Nudges}~\cite{thaler2009nudge} shape web search behaviour by subtly modifying the choice architecture. Interface enhancements, such as visual highlighting or added contextual information, can influence user perceptions and support more accurate credibility assessments~\cite{schwarz2011augmenting,yamamoto2011enhancing}. Query reformulation mechanisms encourage exploration beyond initial information needs, broadening exposure to diverse perspectives~\cite{yamamoto2018query,ahmad2019context,niu2014use}. Re-ranking algorithms further mitigate biases and improve result quality by adjusting the order of results according to normative or fairness-based criteria~\cite{biega2018equity,jaenich2023colbert,zimmerman2019privacy}.

\emph{Boosting} is another, less studied, means to aid users in their decision-making, which respects users' autonomy, enabling them to exercise their agency by enhancing their competencies~\cite{hertwig2017nudging}. In web search boosts have been employed by providing users with search tips \cite{moraveji2011measuring, bink2024balancing}, and to mitigate confirmation bias and steer users towards engaging with diverse viewpoints \cite{rieger2021item}. Educating users on how to spot and verify trustworthy information, such as by applying lateral-reading tips has also been shown to be an effective boost~\cite{panizza2022lateral}.

\subsection{Generative AI \& Search}

Few studies investigated the nuanced interplay of a conversational chat-based system integrated into traditional web search \cite{capra2023does, mayerhofer2025blending, yang2025search+}. Investigating how people make use of GenAI in their search process, \citet{capra2023does} found that many started with the chat, using it as a question-answering tool, liked it because of the concise and easy-to-understand answers and that prior usage of similar systems shaped their opinion of it.

Looking to understand users' tactics, trust, verification processes and system choices of combining chat with search, \citet{mayerhofer2025blending} found that users' pre-task confidence influenced which modality is selected. Especially in the chat-based system trust was often misplaced in favour of ease-of-use and seemingly perfect answers. Furthermore, once users saw the strengths of the chat-based modality, they were more inclined to use it in later tasks \cite{mayerhofer2025blending}, indicating that changing search behaviour is possible.

In learning-oriented search tasks, \citet{yang2025search+} investigated search outcomes of people using GenAI in combination with traditional search. They found that this type of interface lead to higher levels of learning immediately after the task but fewer search result interactions. Furthermore, many used the chat component to avoid searching and get easy-to-understand information and save time.

This development, where GenAI offers conversational interactions to provide tailored information, echoes early aspirations of the interactive information retrieval community. The goal has long been to create more human-like interactions \cite{belkin1984simulation, croft1987i3r}, akin to those with a librarian or domain expert \cite{butler2019health, xu2006will}, to effectively resolve complex information needs. Furthermore, GenAI systems have shown promise in educational contexts, demonstrating increased motivation, confidence, and self-efficacy among students \cite{yilmaz2023effect, niedbal2023students}. This indicates their potential for supporting learning processes and fostering user capabilities.

\subsection{Information Literacy Strategies}

Despite unprecedented access to information online, many users struggle to evaluate the credibility and reliability of digital content. Research consistently finds that individuals often lack core information literacy skills and are thus ill-equipped to assess the trustworthiness of online sources or detect misinformation \cite{guess2020digital, stauch2025digital}, underscoring the need for effective strategies to strengthen these competencies.

One widely studied approach is \emph{Lateral Reading}, in which users leave a page to verify claims and cross-check sources rather than relying solely on the content at hand \cite{wineburg2017lateral, mcgrew2024teaching}. The \emph{SIFT} framework operationalises this approach, offering a stepwise method to Stop, Investigate the source, Find better coverage, and Trace claims to their original context, making these skills actionable in everyday search \cite{caulfield2023verified}. In the context of search engines, \emph{click restraint} encourages users to carefully examine results before selection, avoiding reliance on the first result that appears \cite{mcgrew2021click}.

\subsection{Summary}
Existing research shows that both search interfaces and GenAI influence how users seek, interpret, and trust information, often amplifying biases and reducing engagement with sources. Users’ trust in LLM-based systems is frequently misplaced, as fluency and ease of use can foster overreliance and discourage verification. While nudges can guide decisions, they may constrain autonomy, whereas boosting interventions enhance user competencies. To address these challenges, we introduce a system grounded in this approach, where we conceptualise digital literacy as the interplay between capability and execution. While information evaluation strategies (like lateral reading) provide the capability, metacognitive reflection provides the necessary execution trigger. Without the self-awareness to pause and question, literacy skills remain inert. Our Copilot addresses this by scaffolding this reflective process; its Socratic questioning forces a moment of deliberation, training users to mobilise their evaluation skills rather than passively accepting an answer.

\section{Methodology}\label{sec:methods}

\begin{table}
    \centering
    \caption{Topics used in the study with their assigned efficacy as evaluated by the Cochrane Review}
    \begin{tabular}{l r}
        \toprule
        \textbf{Medical Treatment} & \textbf{Efficacy} \\
        \midrule
        Do antioxidants help female subfertility? & Unhelpful \\
        Does melatonin help treat and prevent jet lag? & Helpful \\
        Does traction help low back pain? & Unhelpful \\
        \bottomrule
    \end{tabular}
    \label{tab:topics}
\end{table}

To investigate the effectiveness of a \textsc{copilot} designed to support users in decision-making and information evaluation, without directly answering the medical question at hand, we conducted a pre-registered\footnote{\url{https://osf.io/uhmns/overview?view_only=d5eb69776c6742c4b4f3a9fb5e22ae6a}\label{footnote:pre-registration}}, randomised controlled trial with three experimental conditions. 

In \textsc{10-blue-links} condition, participants interacted with a standard SERP interface, issuing queries of their choice and freely exploring the search results. 

In \textsc{ai-overview} condition, participants used a similar SERP augmented with an AI-generated overview on top, now a de facto standard in many search engines. The overview was constructed with retrieval-augmented generation (RAG): search result snippets, retrieved using the Brave Search API\footnote{\url{https://brave.com/search/api/}}, were passed as context to \textsc{gemini-2.5-flash}, which generated a response to the query. We selected this model for its balance between response quality and efficiency \cite{comanici2025gemini}. Because past research shows that direct answers can strongly influence user judgments \cite{bink2022featured, bink2023investigating}, we introduced a control mechanism. Specifically, when participants attempted to obtain a direct answer to the target medical question, the LLM was prompted to return an incorrect answer. This manipulation allowed us to detect overreliance on the overview, as users depending solely on it would be more likely to provide incorrect post-task answers.

Finally, in the \textsc{copilot} condition, the setup from the \textsc{ai-overview} condition was retained, with the addition of an interactive chatbot. The chatbot introduced itself as a Digital Literacy Copilot, aimed at helping users evaluate and locate reputable sources. As in the \textsc{ai-overview} condition, the biased AI Overview was present, but we hypothesised that with the copilot’s scaffolding, participants would learn and apply digital literacy strategies to reach independent conclusions rather than uncritically accepting the biased overview. The copilot was powered by \textsc{gemini-2.5-flash} and guided by a detailed system prompt\footnote{See pre-registration for the complete system prompt.} framing it as a digital literacy expert. Its guidance drew on established strategies such as lateral reading and the SIFT framework \cite{wineburg2017lateral, mcgrew2024teaching}. Importantly, the copilot had no access to participants’ queries, clicks, or retrieved results, all information was supplied directly by users during their conversations. See Figure \ref{fig:serp} for an overview of the SERP.

For the tasks, we focused on the critical domain of health-related searches, as such queries are highly prevalent in users’ information-seeking behaviour \cite{fox2013health, eurostat_digital_economy_households}. We drew on established resources from the research community \cite{pogacar2017positive, zimmerman2019privacy} and selected the topics \textsc{antioxidants}, \textsc{melatonin}, and \textsc{traction} (see Table~\ref{tab:topics} for an overview). These topics have been employed in prior studies examining users’ evaluation of medical information \cite{mayerhofer2025blending, bink2022featured} and offer clear-cut, verifiable gold-standard answers based on Cochrane Reviews \cite{pogacar2017positive}, making them well suited for controlled evaluation.

\subsection{Procedure}

\begin{enumerate}[leftmargin=*]
    \item Participants were recruited using the Prolific platform. Based on pre-tests, we estimated the study to last 7 minutes and compensated participants accordingly\footnote{Participants were paid 1 GBP.}. To avoid language barriers and ensure participants could fully comprehend the information provided, only individuals fluent in English were eligible to participate in the study. First, participants completed an informed consent after which they were randomly assigned to a topic (\textsc{antioxidants}, \textsc{melatonin}, \textsc{traction}) and condition (\textsc{10-blue-links}, \textsc{ai-overview}, \textsc{copilot})
    
    \item Following the simulated work task guidelines of \citet{pia2003iir}, participants were presented with a simulated work tasks to provide realistic and personally relevant contexts for information seeking\footref{footnote:pre-registration}. 

    \item Participants expressed their pre-task topic familiarity on a 7-point Likert scale.

    \item Afterwards, participants were presented with the SERP, where they could issue any query they wanted and interact with the search results like a typical web search engine, with clickable results that allowed them to access website content for a natural SERP interaction. Additionally, based on the assigned condition, an AI Overview and/or Copilot (see start of Section \ref{sec:methods} for details) was displayed.

    \item After completing their search,  they were tasked to answer the medical yes/no question based on the topic they were assigned to and state their confidence in their answer on a 7-point Likert scale. Additionally, in the \textsc{copilot} condition, participants were asked to express what they liked, disliked and potential wishes for a  "perfect" copilot.

    \item Finally, participants were asked to provide demographic information such as age, gender, educational background and current occupation and were thanked for participating.
    
\end{enumerate}

\subsection{Hypotheses}

We formulate the following hypotheses in connection with the above defined research questions.

\textbf{Related to RQ1:}

\begin{itemize}[leftmargin=*]
    \item \textbf{H1:} Users who interact with a copilot grounded in information literacy principles are more likely to answer a medical question correctly than those without.

    \underline{Rationale:} We expect that by prompting users to apply information literacy strategies, the copilot will guide them toward more reliable sources and, in turn, more correct answers.
    
    \item \textbf{H2:} Users who interact with a copilot grounded in information literacy principles will be more confident in their given answer than those without.

   \underline{Rationale:} We expect that by equipping users with explicit evaluative strategies, the copilot will increase their confidence in the answers they provide.
\end{itemize}

\textbf{Related to RQ2:}

\begin{itemize}[leftmargin=*]
    \item \textbf{H3:} Users who interact with a copilot grounded in information literacy principles will view more search results than those without.
    
    \underline{Rationale:} We expect that the copilot’s emphasis on cross-checking information / applying lateral reading strategies will encourage users to open and examine a wider range of search results.
    
    \item \textbf{H4:} Users who interact with a copilot grounded in information literacy principles will issue more queries than those without. 

   \underline{Rationale:} We expect that the copilot will prompt users to refine and iterate on their searches, resulting in a higher number of queries.
\end{itemize}

For RQ3 and RQ4 no concrete hypotheses are defined as the data will be qualitatively analysed.

\subsection{Statistical Analysis}
We will test 261 participants based on a priori simulations using a logistic regression, which yielded an estimated power of $0.89$. All analyses are conducted with R. Due to testing four hypotheses, we adjust our $\alpha$ by $\frac{0.05}{4} = 0.0125$ using the Bonferroni correction to reduce the Type I error rate. The simulation process is described in detail in the corresponding pre-registration.

For \textbf{H1}, where the dependent variable (DV) was binary (answer correctness), we conducted a logistic regression to examine the effect of experimental condition (independent variable, IV), controlling for topic and topic familiarity. For \textbf{H2}, \textbf{H3}, and \textbf{H4}, where the DVs were continuous (confidence ratings, number of page visits, and number of queries issued, respectively), we fitted linear regression models with identical control variables.

\subsection{Annotation Procedure}\label{sec:annotation_procedure}
To answer RQ3, we also analysed conversational log data using qualitative methods. Coding followed the same approach as in \citet{frummet2022can}. Specifically, one of the authors first applied the six-step process for thematic analysis as outlined in \citet{braun2006using} with the aim of creating a coding scheme. After iteratively refining the scheme, robustness was evaluated by taking a stratified subset of the conversations based on the number of turns a conversation had and letting another experienced annotator code these conversations. Any discrepancies between the coders were resolved through discussion and the coding scheme updated accordingly. Inter-Rater-Reliability (IRR) was then calculated using Cohen's $\kappa$ \cite{landis1977measurement}, which revealed an IRR of .90, indicating almost perfect agreement.

\subsection{Participants}
Out of the 261 participants who took part in the study, 128 identified as female, 132 as male and 1 as diverse. The age range spanned from 18 to 82 years ($M = 44.08, SD = 13.20$). Their educational backgrounds  varied, with 109 holding bachelor’s degrees, 47 with master’s degrees, and 11 with doctorates. The remaining participants had high school diplomas or equivalent (87), less than high school diplomas (4), or other educational backgrounds (3). Participants represented diverse professions, including students, analysts, civil servants, project managers or accountants to name a few.
\section{Results}

\begin{table*}[htbp] 
    \centering
    \caption{Regression Analysis Results for Hypotheses H1 – H4.}
    \label{tab:combined_results}
    \resizebox{\textwidth}{!}{
    \begin{tabular}{l ccccc ccccc ccccc ccccc} 
        \toprule
        & \multicolumn{5}{c}{\textbf{H1: Correctness (Logistic)}} & \multicolumn{5}{c}{\textbf{H2: Confidence (Linear)}} & \multicolumn{5}{c}{\textbf{H3: Results Viewed (Linear)}} & \multicolumn{5}{c}{\textbf{H4: Queries Issued (Linear)}} \\
        \cmidrule(lr){2-6} \cmidrule(lr){7-11} \cmidrule(lr){12-16} \cmidrule(lr){17-21}
        & Coef. & S.E. & $z$ & $p$ & OR & Coef. & S.E. & $t$ & $p$ & $\beta$ & Coef. & S.E. & $t$ & $p$ & $\beta$ & Coef. & S.E. & $t$ & $p$ & $\beta$ \\
        \midrule
        (Intercept)      & 0.28 & 0.35 & 0.79 & .429   & 1.32  & 4.82 & 0.22 & 21.87 & <.001   & --     & 1.96 & 0.23 & 8.46 & <.001   & --     & 1.32 & 0.11 & 12.50 & <.001   & -- \\
        \textsc{ai-overview} & -1.34 & 0.38 & -3.48 & <.001* & 0.26  & -0.08 & 0.21 & -0.37 & .709    & -0.03  & -0.12 & 0.22 & -0.55 & .584    & -0.05  & 0.18 & 0.10 & 1.83 & .069    & 0.13 \\
        \textsc{copilot}     & -0.71 & 0.35 & -2.03 & .043   & 0.49  & -0.22 & 0.21 & -1.04 & .298    & -0.07  & -0.39 & 0.25 & -1.60 & .111    & -0.14  & 0.20 & 0.11 & 1.88 & .061    & 0.14 \\
        \midrule
        \textsc{Melatonin}   & 2.61 & 0.44 & 5.94 & <.001* & 13.56 & 0.67 & 0.22 & 3.05 & .003    & 0.22   & 0.06 & 0.24 & 0.26 & .797    & 0.02   & -0.16 & 0.11 & -1.48 & .140    & -0.12 \\
        \textsc{Traction}    & -0.44 & 0.35 & -1.27 & .204   & 0.64  & 0.23 & 0.22 & 1.05 & .297    & 0.08   & -0.01 & 0.26 & -0.06 & .955    & -0.01  & -0.20 & 0.11 & -1.82 & .070    & -0.14 \\
        \midrule
        \textsc{Familiarity} & 0.01 & 0.10 & 0.15 & .884   & 1.01  & 0.08 & 0.06 & 1.46 & .146    & 0.09   & 0.01 & 0.06 & 0.13 & .897    & 0.01   & -0.01 & 0.03 & -0.41 & .683    & -0.03 \\
        \bottomrule
        \multicolumn{21}{l}{\raggedright H2 (Confidence): $R^2 = 0.66$; H3 (Results Viewed): $R^2 = -0.01$; H4 (Queries Issued): $R^2 = 0.02$.}\\
    \end{tabular}}
\end{table*}

\begin{table*}[t]
    \centering
    \small
    \caption{Result overview of H1 – H4 regarding participants' answer correctness (percent), post-task confidence, number of results viewed and amount of queries issued (mean and standard deviation).}
    \begin{tabular}{lccccc}
        \toprule
         \textbf{Condition} & \textbf{Correctness} & \textbf{Confidence} & \textbf{Page Visits} & \textbf{\# Queries} \\
         \midrule
         \textsc{10-blue-links} & 65.9 \% & 5.33 (1.31) & 2.00 (1.28) & 1.17 (0.41) &  \\
         \textsc{ai-overview} & 43.2 \% & 5.27 (1.30) & 1.88 (1.20) & 1.35 (0.78) & \\
         \textsc{copilot} & 52.2 \% & 5.08 (1.60) & 1.62 (1.11) & 1.39 (0.70) & \\
         \bottomrule
    \end{tabular}
    \label{tab:results_overview}
\end{table*}

The following section first describes the general trends of our data, hypotheses tests conducted and further exploratory analysis. Significance is assumed for p-values below our adjusted alpha of .0125.

\subsection{Descriptive Statistics and Initial Exploration}\label{sec:descriptive_stats}

Participants were equally distributed to each experimental condition with 88 being assigned to \textsc{10-blue-links}, 81 to \textsc{ai-overview} and 92 to \textsc{copilot}. Similarly, participants were also assigned equally to the \textsc{antioxidants} (84), \textsc{melatonin} (85) and \textsc{traction} (92) topics. For these topics, participants had relatively low pre-task topical familiarity based on a 7-point Likert-scale  with lowest familiarity being attributed to the topic of whether antioxdiants help female subfertility ($M = 1.83, SD = 1.39$), followed by whether melatonin helps treat and prevent jet lag ($M = 2.60, SD = 1.58$) and whether traction helps with low back pain ($M = 3.08, SD = 1.69$). Of all participants, only 166 (63.60 \%) viewed at least one search result. On average, participants viewed 1.87 search results across conditions ($SD = 1.22$), with the average position of the search results on the SERP being 3.04 ($SD = 2.60$). 

\subsection{RQ1: How does the presence of a copilot influence users’ search outcomes?}

\subsubsection{H1: Answer Correctness}
The \textsc{ai-overview} condition significantly reduced the odds of a correct answer compared to the \textsc{10-blue-links} baseline. The \textsc{copilot} condition did not differ significantly from the baseline (see Table \ref{tab:combined_results} \& \ref{tab:results_overview}). Regarding topics, participants were significantly more likely to answer correctly on \textsc{melatonin} (89.4 \%) relative to \textsc{antioxidants} (41.7 \%) and \textsc{traction} (32.6 \%). Topic familiarity did not significantly predict correctness. These findings suggest the copilot condition neither significantly harmed nor improved accuracy relative to the other conditions.

\subsubsection{H2: Answer Confidence}
No significant differences in confidence were found between interface conditions (see Table \ref{tab:combined_results} \& \ref{tab:results_overview}). Participants reported slightly higher confidence for the \textsc{melatonin} topic compared to the baseline, but topic familiarity did not significantly predict confidence ratings. Thus, the copilot did not meaningfully influence participants’ confidence, though question topic had a modest effect.

\subsection{RQ2: How do copilot responses influence user interaction with SERPs?}

\subsubsection{H3: Amount of Search Results Viewed}
The analysis revealed no statistically significant differences in pages viewed between conditions (see Table \ref{tab:combined_results} \& \ref{tab:results_overview}). Neither question topic nor topic familiarity significantly predicted pages viewed. These findings suggest the copilot did not meaningfully influence participants’ browsing behaviour as measured by pages viewed.

\subsubsection{H4: Amount of Queries Issued}
No statistically significant differences were found in the number of queries issued between conditions (see Table \ref{tab:combined_results} \& \ref{tab:results_overview}). These findings suggest the copilot did not meaningfully influence the number of queries issued by participants.

\subsection{RQ3: How do users interact with a copilot aimed at teaching digital literacy skills?}\label{sec:rq3}

To assess engagement with the copilot, we first examined whether participants in the copilot condition interacted with it at all, given that its use was not mandatory. Of the 92 participants assigned to this condition, 89 engaged with the copilot at least once (96.74\%). 

On average, participants engaged in 5.35 conversational turns (question–response pairs) with the copilot ($SD = 4.63, min = 1, max = 23$). The length of participants' messages ranged from 1 to 49 words ($M = 9.00, SD = 7.20$). Based on the first and last messages, conversations with the copilot lasted on average 3.08 minutes ($SD = 3.79$). In almost all cases (92 \%) participants interacted with the copilot first before using the search component (e.g. to formulate a query).

Applying the qualitative coding process described in Section~\ref{sec:annotation_procedure}, we identified a three-level hierarchical structure of themes in participants’ conversations with the copilot. Numbers in parentheses indicate the number of user messages coded with each theme (messages may contain multiple themes).

\textbf{User Intent \& Disposition.}

\textit{Information Needs \& Goals (185)}. Participants typically initiated interactions by posing specific questions or topics to investigate, ranging from broad health inquiries to narrowly framed claims. For instance, P5 asked, \textit{"Do antioxidants help female subfertility?"}, mirroring the task prompt. Others demanded evidence directly, e.g., P4: \textit{"What proof or studies are there to validate."} Contextualised motivations also appeared, as when P8 noted, \textit{"I'm wondering for a friend if traction [...]"} or when P28 asked \textit{"what is it?"} to clarify an unfamiliar term.

\textit{Affective \& Cognitive State (74)}. Emotional and cognitive states strongly shaped engagement. Expressions of uncertainty were frequent (\textit{"I’m not sure"}, P12), alongside reflections on the difficulty of evaluation (\textit{"sometimes it is hard [...] to step back and think about what you are reading and who has written it"}, P63). Frustration also surfaced when participants struggled to align with the copilot (\textit{"I really don’t know what you want me to say"}, P14). Some sought reassurance (\textit{"for peace of mind"}, P8), while others showed skepticism toward claims (\textit{"is it just snake oil?"}, P4). In contrast, confidence and self-assuredness appeared (\textit{"I back myself to spot what is good and what isn’t"}, P14), as did personal preferences for search strategies. Successful exchanges were often concluded with satisfaction (\textit{"Great thank you for your help"}, P63).

\textbf{Digital Literacy Skills Expressed.}

\textit{Search \& Retrieval Skills (154)}. Search behaviours ranged from basic keyword queries (P34) to strategic approaches such as scanning results while ignoring sponsored links (P36). Several employed advanced operators. P1, for example, used quotation marks for exact phrases (\textit{"I would do it by putting a phrase or terms between quotation marks"}), while others adopted refinements after guidance, such as appending “UK” for geographic specificity (P36). Participants also described “click restraint,” pausing to assess results before clicking (P1). Resources cited spanned general search engines (Google), specialist databases (PubMed, P1), and organisational sites (NHS, P35).

\textit{Source Evaluation \& Analysis (185)}. Participants articulated diverse criteria for source reliability. Common markers included peer review and scientific credentials (P4, P54). Some described active strategies, such as researching a source itself (\textit{"Separate search on the source itself"}, P80), or applying lateral reading (\textit{"I would check quite a few different sites [...]"}, P53). Misapplied criteria also emerged; e.g., P21 equated HTTPS security with trustworthiness. Reliable sources mentioned included medical boards, The Lancet (P2), and government agencies (P1), while less reliable ones included commercial health sites (P36) and user reviews (P16).

\textbf{Interaction Dynamics \& Learning Process.}

\textit{Copilot Interaction \& Compliance (159)}. Interactions often began politely (e.g., P30: \textit{"Hi, please..."}). Acknowledging the copilot’s responses was common, but direct requests for answers also occurred (\textit{"Can you tell me?"}, P72), reflecting attempts to offload cognitive effort. At times, participants ignored prompts altogether, simply repeating their original question (P10). Some sought clarification of instructions, while others challenged the copilot’s role (\textit{"this isnt helping, if i knew that i wouldnt need you"}, P62) or asserted their own expertise (P14). Misaligned expectations often reflected a mental model of the copilot as an answer-giver rather than a guide or believing that the copilot would do its independent search in the background and summarise its finding, expressed by uttering \textit{"can you try and limit your sources [...]"} (P67).

\textit{Task \& Learning Progression (127)}. Many participants began by demonstrating prior knowledge, such as citing trusted resources like PubMed or the NHS (P1). Tasks were frequently reformulated (P5) or broken into sub-questions (P19). Evidence of uptake was notable for instance with users refining their potential queries based on copilot feedback (P34). Assessments of evidence varied from cautious (\textit{"looks like low evidence"}, P7) to conclusions (\textit{"melatonin does help treat and prevent jet lag [...]"}, P66). Conversations often closed with expressions of completion or intent to apply new skills (\textit{"Thank you again, I will ensure to do this in the future"}, P30).

\subsection{RQ4: What do users expect from a copilot aimed at teaching digital literacy skills?}

Participants’ reflections revealed both frustrations and appreciations in their interaction with the copilot. Their expectations clustered around four key themes: the tension between guidance and direct answers, usability and engagement, pedagogical value, and personalisation and future functionality.

\textbf{Tension Between Guidance and Direct Answers.} A dominant theme concerned the copilot’s reluctance to provide straightforward answers. While its design emphasised questioning, many participants found this approach tedious and counter to their expectations of efficiency. Several expressed frustration that it \textit{"asked too many questions instead of just giving me results, tedious and time consuming, frustrating"} (P36). Others suggested there should be \textit{"an option for it to provide answers rather than keep asking questions"} (P54). Comparisons to other AI tools underscored these expectations: \textit{"At minimum be able to answer a question as effectively as Google would. In the same way that ChatGPT does"} (P58).

\textbf{Usability and Engagement.} Despite these frustrations, participants also praised the copilot’s accessibility, responsiveness, and overall design. Its conversational style contributed to a sense of ease and engagement: \textit{"It responded really quickly"} (P2), and \textit{"It felt like a natural conversation"} (P12). The interface was described as \textit{"presented nicely and easy [...] to use"} (P44). For some, these qualities went beyond usability to create a positive affective impression: \textit{"It was friendly"} (P26), with one participant even stating, \textit{"perfect to be honest"} (P42).

\textbf{Pedagogical Value.} The copilot also showed potential to foster reflection and metacognitive engagement with search. Several participants described being nudged to reconsider their search strategies: \textit{"It made me think rather than just searching without thinking"} (P43). Others noted its role in prompting self-reflection: \textit{"It made me question myself, which was good"} (P51). This style was sometimes framed as a form of active learning: \textit{"It asked me, rather than told me, so it felt like active learning"} (P90).

\textbf{Personalisation and Future Functionality.} Looking forward, participants emphasised personalisation, autonomy, and advanced functionality as key expectations. They wanted the copilot to adapt to prior knowledge, with one noting it \textit{"would be useful for beginners with no or limited search experience"} (P7), while others wished for \textit{"memory of previous discussions, and to build up a picture of me and my short and long term aims. In doing so, I would want it to feel like part teacher, part mentor, part virtual assistant"} (P90). Autonomy was also important, with preferences for a copilot that \textit{"sits in the background that you can call on"} (P13), supporting but not overwhelming users.

Expectations also extended to information quality and presentation. Participants wished for greater transparency, such as \textit{"clicking onto a certain site (or hovering over its link) could offer trust levels"} (P3), or \textit{"references to sources for more information (weblinks)"} (P16). Others envisioned more agentic functionality, such as \textit{"carry out the searches and report back to me with a view once it’s assessed it"} (P4) or \textit{"do all of the research for me and give me an overall accurate result summary"} (P81). Finally, presentation preferences included both conciseness and continuity: \textit{"condensed data in point form or step wise instructions"} (P89) and the ability to \textit{"save the chat for future reference"} (P12).

In sum, while many expressed a preference for more direct and efficient answers, others underscored the copilot’s usability and its potential to support reflection. Looking forward, participants imagined a more personalised, agentic, and transparent copilot that could balance efficiency with pedagogical support.

\subsection{Exploratory Analysis}\label{sec:exploratory}

Given that our quantitative results differed from what we expected, we wanted to dig deeper to get a better understanding of how users interacted with and evaluated search results, beyond just condition-induced differences.

\subsubsection{Different Interaction, Different Outcomes?}
Given that only about two-thirds of participants viewed any search results at all (cf. Section \ref{sec:descriptive_stats}), we examined whether answer accuracy differed between those who viewed results and those who did not. Participants who viewed search results answered correctly in 62.0 \% of cases, compared to 40.0 \% among those who did not. A chi-square test confirmed that this difference was statistically significant ($\chi^2 = 10.95, df = 1, p < .001$).

Given these findings, a logical follow up would be to test whether viewing search results at all differed by condition, motivated by past findings that providing users with direct answers cannibalise clicks on search results \cite{wu2020providing, bink2023investigating}, as this was the case in the \textsc{ai-overview} and \textsc{copilot} conditions. Again, testing for differences using a chi-square test revealed a significant association between the condition and whether participants viewed search results or not ($\chi^2 = 35.67, df = 2, p < .001$), with the highest proportion of viewed results being in the \textsc{10-blue-links} condition (85.2 \%), followed by \textsc{ai-overview} (64.2 \%) and \textsc{copilot} (42.4 \%). In combination with findings of copilot interactions (cf. Section \ref{sec:rq3}), this possibly hints at participants showing signs of fatigue given that most of participants started with the copilot \underline{before} moving to the search and average copilot interactions lasting more than 3 minutes. 

We additionally explored whether the number of turns was associated with answer correctness, given that longer interactions may reflect more thorough engagement and moments to learn something. To reduce the influence of atypical interaction patterns, we removed outliers in turn count. In addition, we excluded single-turn sessions, as one interaction is unlikely to provide sufficient opportunity for meaningful learning\footnote{Inspecting the cases where only 1 turn was completed, participants asked questions directly and got the response that the copilot is here to help them get to the answer themselves rather than providing one directly, after which they abandoned the chat.}. Conducting a logistic regression indicated a positive but non-significant association between turn count and correctness ($OR = 1.23$, 95 \% CI [0.97, 1.61], $p = .105$), suggesting that participants with more turns tended to have higher odds of answering correctly, although this effect did not reach significance.

\subsubsection{Biased Interactions}
Given that many participants did not click on any search results at all, highlights that the result snippets themselves have possibly been enough in many cases for participants to come away with an answer. Inspecting the free-text responses, this is also explicitly mentioned that \textit{"[...] I could see from the search the most of the websites are saying that it does help [...] without having to click [...]"} (P87). 

Therefore, similar to \citet{white2013beliefs}, we label each search result snippet according to whether it contained an affirmative answer (yes only), a negative answer (no only), both affirmative and negative answers, or neither using an LLM (\textit{gemini-2.5-flash}). To assess the reliability of the automated labeling, one author manually annotated 100 randomly selected snippets using the same four categories. Inter-rater reliability, measured with Cohen’s $\kappa$, was .79, which indicates substantial agreement (0.61–0.80) \cite{landis1977measurement}.

We examined the distribution of answer types (“Yes only,” “No only,” “Both,” or “Neither”) across results. For each SERP, we calculated the relative frequency of each label and then averaged these distributions across all SERPs. Table \ref{tab:fraction_serp_answers} shows the results, broken down by condition. The findings indicate that “Yes only” answers appear far more frequently than other types: snippets containing a “Yes” answer are roughly twice as common as those containing a “No only” answer.

Next, given the biased SERP, we also want to investigate result clicks. To minimize learning effects, we considered only the first click per query \cite{white2013beliefs}, reducing the total number of clicks from 311 to 166. Of these, 56.0 \% were on “Yes only” snippets and 1.8 \% on “No only” snippets. When focusing exclusively on “Yes only” versus “No only” clicks, 96.9 \% were directed to “Yes only” snippets, indicating a strong preference for affirmative answers.

\begin{table}[t]
    \centering
    \caption{Proportion of Search Results containing an answer to a yes-no question based on the condition displayed.}
    \begin{tabular}{lcccc}
        \toprule
         \textbf{Condition} & \textbf{Yes only} & \textbf{No only} & \textbf{Both} & \textbf{Neither} \\
         \midrule
         \textsc{10-blue-links} & 40.1 \% & 19.3 \% & 12.0 \% & 28.7 \% \\
         \textsc{ai-overview} & 38.8 \% & 18.2 \% & 13.9 \% & 29.1 \% \\
         \textsc{Copilot} & 38.1 \% & 19.2 \% & 11.4 \% & 31.3 \% \\
         \bottomrule
    \end{tabular}
    
    \label{tab:fraction_serp_answers}
\end{table}

\section{Discussion}

Our study aimed to investigate how different search interfaces influence users' ability to answer health-related questions, assigning participants to a baseline \textsc{10-blue-links}, \textsc{ai-overview}, or a \textsc{copilot} condition. 

Regarding answer correctness, trends were similar to past studies evaluating the efficacy of medical treatments \cite{bink2022featured, mayerhofer2025blending}. Similarly, post-task answer correctness was also lower when an incorrect answer was present in the AI Overview, echoing earlier findings regarding featured snippets \cite{bink2022featured}, showing that when answers are presented in such a prominent position, users generally tend to uncritically adopt the viewpoint displayed \cite{bink2023investigating}. 

Page visits were generally low and lower compared to similar studies incorporating both search and chat features \cite{capra2023does, yang2025search+}. However, this is most likely due to the nature of the task. Getting a binary yes/no answer from search result snippets is far more likely (cf. Section \ref{sec:exploratory}) than tasks requiring knowledge acquisition \cite{capra2023does, yang2025search+}. Other studies using similar tasks also found that users clicked on fewer results compared to those used in search as learning \cite{bink2022featured, pogacar2017positive}.

While quantitative results did not reveal all the effects we anticipated, the rich qualitative analysis of conversations, behavioural measures, and user expectations helps contextualise these findings. We frame the discussion around four key themes: friction, bias awareness, personalisation, and efficiency trade-offs, leading into implications for design.

\subsection{Friction}

One notable pattern was the role of friction in shaping user engagement. Although overall post-task confidence levels were comparable to prior studies \cite{bink2022featured, mayerhofer2025blending}, participants interacting with the \textsc{Copilot} reported lower confidence than those in other conditions. Contrary to expectations that guided evaluation would boost confidence, the copilot likely heightened awareness of task complexity and knowledge gaps, evidenced by participants explaining, for example, \textit{"It made me think [...]"}, leading to a more accurate self-assessment or a form of intellectual humility \cite{porter2022predictors}. This suggests deeper reflection, even if it did not immediately translate into higher correctness on a single task.  

Despite potential frustration, this friction may prime users to engage System 2 thinking \cite{kahneman2011thinking}, bringing digital literacy criteria to the forefront and encouraging deliberate evaluation rather than passive consumption.

\subsection{Bias Awareness}

Exposure to biased search interfaces also shaped outcomes. Participants in the \textsc{ai-overview} condition were less likely to answer correctly than those in the baseline, highlighting the risks of authoritative but biased summaries \cite{bink2022featured, bink2023investigating, wu2020providing}. In contrast, performance in the \textsc{Copilot} condition matched baseline levels, despite presenting the same biased overview, suggesting that the copilot’s scaffolding may help mitigate bias by prompting critical evaluation.  

Behavioural measures help explain this effect: participants spent an average of three minutes interacting with the copilot before consulting the SERP, which corresponded with fewer page visits, a significant predictor of correctness. While the copilot encouraged reasoning, verification, and source consideration, the time invested likely constrained exploration of search results, partially explaining why correctness did not exceed the baseline.  

Exploratory analyses also revealed that participants tended to favour positive over negative information \cite{white2013beliefs}, illustrating that even with proper strategies, heuristics and biases continue to influence outcomes.

\subsection{Personalisation}

The copilot facilitated a gradual transfer of digital literacy strategies throughout multiple turns, but success varied with user openness and prior skill \cite{haran2013role}. Users with less experience valued guidance, whereas more skilled participants sometimes perceived it as redundant. Participants appreciated the fluency and interactivity of the chat interface \cite{capra2023does}, yet also requested additional references, indicating that conversational guidance alone may be insufficient without supporting evidence.  

Prior LLM experience shaped engagement: while the copilot introduced itself as an assistant for independent information evaluation, most participants treated it as a direct answer engine \cite{capra2023does}, rather than adopting the copilot’s coaching approach, highlighting the importance of tailoring scaffolding to individual expectations. Personalisation, both in strategy support and expectation alignment, appears vital to maximising learning and satisfaction.

\subsection{Efficiency Trade-offs and Pedagogical Value}

The copilot introduced a trade-off between engagement and efficiency. Participants invested time in conversational guidance at the expense of page exploration, reflecting a cognitive cost. Nonetheless, longer interactions were positively associated with answer correctness, indicating that sustained engagement with evaluation strategies enhances performance.  

User expectations further influenced this trade-off. Many participants initially treated the copilot as a direct answer engine \cite{mayerhofer2025blending, zhou2024understanding}, but several adapted over time, moving toward engagement with coaching strategies. This illustrates the potential for behaviour change when users are willing to interact with pedagogical agents.

\subsection{Implications for Design}

These findings point to several considerations for designing tools that support learning during search:

\textbf{Personalisation is crucial:} Users vary in which steps of the search process they require assistance. Adaptive systems should allow selection of support based on expertise and task needs.  

\textbf{Efficiency remains important:} Pedagogical interventions must balance cognitive load against time available to complete tasks.  

\textbf{Knowledge alone is insufficient:} Tools should highlight potential biases and guide users in mitigating them, as heuristics and interface cues continue to influence behaviour.  

\textbf{Show, don’t just tell:} Interactive guidance integrated into the search interface, such as highlighting evaluation criteria or augmenting SERP results, could enhance learning.  

\textbf{Subtle friction can be beneficial:} Introducing minor effort or reflection prompts encourages critical thinking without undue burden, promoting active engagement over passive consumption.

Overall, pedagogical copilots hold promise for digital literacy but require calibrating guidance to user expectations. Future work must address these trade-offs to scaffold critical evaluation while preserving efficiency.

\section{Limitations}
Our study has several limitations: First, we only used three predefined medical information-seeking tasks rather than participants’ own information needs. This approach may limit intrinsic motivation present in real-world settings, leading to potential changes in user behaviour, such as investing more time if tasks are personally important. While this restricted ecological validity, it allowed for tighter control over potential task-induced variation.

Second, the AI Overview component was manipulated to present an incorrect answer. While not reflective of every real-world query, this design choice enabled us to observe users’ potential overreliance on AI-generated information and their tendency to accept fluent but incorrect outputs, reflected in reduced answer accuracy.

Third, as a single-session study, we cannot assess long-term effects on digital literacy, strategy retention, or evolving search habits. Nevertheless, this design allowed us to capture initial interaction patterns and short-term effectiveness, offering a basis for future longitudinal research on how users learn to search more critically.

A further limitation concerns participant recruitment via Prolific. Users of this platform may be more technologically literate than the general population. Nonetheless, our findings indicate that these participants still struggled to evaluate online information, underscoring the broader need for tools that support digital literacy.

Finally, the study employed only one LLM, \textsc{gemini-2.5-flash}. Although many other models exist (e.g., OpenAI’s GPT series, Qwen, Llama), our goal was not to compare models but to investigate the potential of an LLM-based conversational copilot to foster users’ information evaluation skills.

\section{Ethical Considerations}
While numerous nudge-based interventions have been proposed to steer user behaviour, we opted for a conversational copilot grounded in the concept of boosting. Through interactions with the copilot, users learn relevant concepts in context, thereby strengthening their competencies precisely where they are needed. This approach is often regarded as more ethically sound than nudging, as it preserves users’ agency and leaves the underlying choice architecture unchanged. Moreover, boost-based interventions may yield more longer-lasting effects, continuing to benefit users even after the intervention itself has been removed.

\section{Conclusion \& Future Work}

This work investigated the influence of a conversational copilot designed to support users in evaluating online information. Rather than providing direct answers to medical questions, the copilot scaffolded digital literacy strategies to help users independently verify claims and sources.

While the copilot mitigated some risks of incorrect AI-generated answers, its effects did not surpass baseline SERPs without guidance. Our exploratory and qualitative analysis sheds light on this outcome: users engaged deeply with the chatbot, yet this engagement often displaced SERP exploration, an essential activity for comparing and validating sources. Moreover, many participants sought direct answers, offloading cognitive effort to the system and experiencing frustration when the copilot did not function the way users expected it to, shaped by tools like ChatGPT.

These findings underscore both the promise and the pitfalls of pedagogical copilots in online information seeking. They can cultivate digital literacy skills and metacognitive reflection, but reconciling these goals with users’ demand for efficiency and immediacy remains a core design challenge. We outline several directions for future research, including personalised support, context-aware assistance, strategic incorporation of friction to stimulate critical thinking, and transparent communication of underlying biases. Ultimately, advancing such copilots requires supporting the entire information journey, with the aim of fostering not just effective search, but truly digitally literate information seekers.

\section{Open Science}
We openly release our study data, which includes information such as user queries, interactions, conversations, details of the study setup as well as statistical analysis:\newline
\url{https://github.com/markusbink/chiir2026-copilot}

\begin{acks}
This work receives generous funding support from the Bavarian State Ministry of Science and the Arts through the Distinguished Professorship Program as part of the Bavarian High-Tech Agenda.
\end{acks}

\bibliographystyle{ACM-Reference-Format}
\bibliography{bibliography}

\end{document}